\documentstyle[epsf]{article}

\newcommand{\tr}{\mathop{\rm tr}\nolimits}

\newcommand{\Ker}{\mathop{\rm Ker}\nolimits}
\newcommand{\im}{\mathop{\rm Im}\nolimits}
\newcommand{\mod}{\mathop{\rm \; mod \;}\nolimits}

\begin{document}

\title{Reduction of XXZ model with generalized periodic boundary conditions}
\author{A.\,A.\,Belavin\/\thanks{belavin@itp.ac.ru},
        S.\,Yu.\,Gubanov\/\thanks{gubanov@itp.ac.ru} \\
{\it Landau Institute for Theoretical Physics, RAS,}\\
{\it Chernogolovka, Moscow region, 142432, Russia}}

\date{February 5, 2002}

\maketitle

We examine the XXZ model with generalized periodic boundary
conditions and identify conditions for the truncation of the
functional fusion relations of the transfer matrix fusion.
After the truncation, the fusion relations become a closed
system of functional equations. The energy spectrum can  be
obtained by solving these equations. We obtain the explicit
form of the Hamiltonian eigenvalues for the special case where
the anisotropy parameter $q^4=-1.$

$$
$$
Published in {\it Theoretical and Matheatical Physics, Vol. 129, No. 2, pp. 1484-1493, 2001}\\
Translated from {\it Teoreticheskaya i Matematicheskaya Fizika, Vol. 129, No. 2, pp 207-218,
November, 2001}\\
Results published in {\it JETP Letters, Vol. 73, Issue 9, pp. 498-502, May 10, 2001}

\section{Introduction}
The Hamiltonian of the anisotropic Heisenberg-Ising model,
or the XXZ model, has the form~\cite{Goden}
\begin{equation}
H_{XXZ}= \sum_{n=1}^{N} \left( \sigma^{+}_n \sigma^{-}_{n+1}+
\sigma^{-}_n \sigma^{+}_{n+1}
+\frac{q+q^{-1}}{4} \; \sigma^z_n \sigma^z_{n+1} \right).
\end{equation}
We focus on the case where the module of the anisotropy parameter
$q$ is equal to $1$, i.e., $q=e^{i\eta}$.
We consider the generalized periodic ({\it twisted}) boundary conditions
\begin{eqnarray}
\sigma^{\pm}_{N+1}&=& q^{\pm 2 \beta}\sigma^{\pm}_1\\
\nonumber
\sigma^z_{N+1}&=&\sigma^z_1.
\end{eqnarray}
The spin projection
$S^z=\frac12 \sigma^z_1+\ldots+\frac12 \sigma^z_N$,
on the $z$ axis is conserved, i.e.,
$[S^z, H_{XXZ}]=0$.
As established in~\cite{PS}, the energies of chains with different
twist parameters are interconnected such that the energy spectrum
in the chain with the twist parameter $\beta$ from the sector with
$S^z=\beta-1$ contains the energy spectrum of the chain with the
twist parameter $\beta-n$ from the sector $S^z=\beta-1+n$,
where $n$ is an integer, i.e.,
\begin{equation}
\label{ENERG}
E^{(\beta)}_{S^z=\beta-1}=E^{(\beta-n)}_{S^z=\beta-1+n}.
\end{equation}
This properly follows from the quantum group symmetry ${\rm U_q(sl(2))}$.
We say that "one spectrum contains the other" because the
number of vectors in the sector $S^z=\beta-1+n$ is smaller than in
$S^z=\beta-1$.

The XXZ model is connected with the two dimensional classical statistical
lattice six vertex model (the ice model)~\cite{Goden}.
The transfer matrix $\hat{t}_{1/2}(u)$ of the six vertex model commutes
with the Hamiltonian, $[\hat{t}_{1/2}(u), H_{XXZ}]=0$, for all values of
the spectral parameter $u$. This matrix also commutes with itself,
$[\hat{t}_{1/2}(u), \hat{t}_{1/2}(v)]=0$, and is therefore a generating
function of the commuting integrals of motion that are in involution.
In particular, the Hamiltonian itself can be expressed through the
logarithmic derivative of $\hat{t}_{1/2}(u)$ as
\begin{equation}
\label{SVYAZ}
H_{XXZ} = -\frac{N}{2} \cos(\eta) + \sin(\eta)
\frac{d}{d u} \log \hat{t}_{1/2}(u)|_{u=0}.
\end{equation}
Formula~(\ref{SVYAZ}) allows obtain the energy spectrum of the XXZ chain
if the transfer matrix eigenvalues are known.

There exists an infinite family of commuting transfer matrices
$t_{j}(u)$, where $j=0, \frac12, 1, \frac32, \ldots$.
These matrices satisfy the infinite system of recursive functional
fusion relations~\cite{Zhou, KR, YBAT}
\begin{eqnarray}
\label{FUSION}
\nonumber
\hat{t}_{1/2}(u-(j+1/2)\eta) \; \hat{t}_j(u) & = &
t_0(u-(j+1)\eta) \; \hat{t}_{j-1/2}(u+\eta/2) \\
& + & t_0(u-j\eta) \; \hat{t}_{j+1/2}(u-\eta/2),
\end{eqnarray}
where $t_0(u) = \sin^N(u+\eta/2)$.
In this article, we prove that in the case where the anizotropy
parameter is the root of unity, $q^{p+1}=-1$, the transfer matrix
in the representations with the spin $j=p/2$ has zero eigenvalues
on the Bethe eigenvectors if the twist parameter $\beta$ and the
sector $S^z=s$ satisfy the conditions
\begin{equation}
\label{USLOV}
(-1)^{2 \beta} = (-1)^N, \qquad |s| < \min (\beta, \; p+1-\beta)
\mod (p+1)
\end{equation}
Once the eigenvalue of $t_{p/2}(u)$ vanishes, the infinite system
of functional fusion relations~(\ref{FUSION}) is truncated and becomes
a functional equation with respect to $t_{1/2}(u)$. Solving this
functional equation, it is  possible to obtain the $XXZ$ energy
spectrum through formula~(\ref{SVYAZ}).
We let the symbol $V_{p, \beta-s}$ denote the set of the Bethe
eigenvectors on which $t_{p/2}(u)$ has zero eigenvalues.
In this article, $V_{p, \beta-s}$ is fully determined in the terms
of the generators of the quantum algebra  ${\rm U_q(sl(2))}$.
We call the $XXZ$ model reduced onto the set $V_{p, \beta-s}$ the
reduced $XXZ$ model. The energy spectrum of the reduced $XXZ$ model
can be obtained by solving the functional equations for the transfer
matrix of the siz-vertex model.

In Sec. 2, we prove the existence of the Bethe vectors on which
the functional fusion relations are truncated if conditions~(\ref{USLOV})
are satisfied.   All such vectors are identified in Sec. 3, where we define
$V_{p, \beta-s}$ in terms of the generators of the quantum algebra
${\rm U_q(sl(2))}$. In Appendix~A, we derive the formula for the transfer
matrix $t_{p/2}(u)$. The transfer matrix eigenvalues and the energy spectrum
of the reduced $XXZ$ model for the case where the anisotropy parameter
$q^4$ is equal to $-1$ are explicitly calculated in Appendix~B.

\section{Zero eigenvalues}
We prove the existence of zero eigenvalues of the transfer matrix
$t_{p/2}(u)$ assuming that $q^{p+1}=-1$ and conditions~(\ref{USLOV})
are satisfied. The transfer matrix of the siz-vertex model corresponding
to the $XXZ$ model Hamiltonian with twisted boundary conditions has the form
\begin{equation}
\hat{t}_{1/2}(u) = \tr \left( q^{-\beta \sigma^z} R_N(u)\ldots
R_1(u)\right),
\end{equation}
where the  $R$-matrix is
\begin{equation}
R_n(u) =
\left(
\begin{array}{cccc}
\sin(u+\frac{1}{2}(1+\sigma^z_n)\eta) & \sin(\eta) \sigma^{-}_n \\
\sin(\eta) \sigma^{+}_n & \sin(u+\frac{1}{2}(1-\sigma^z_n)\eta) \\
\end{array}
\right).
\end{equation}
The well-known procedure of the algebraic Bethe anzatz (quantum inverse
scattering method)~\cite{TaFa} leads to the transfer matrix eigenvalues
expressed through the roots $\{v_m \}$ of the Bethe equations,
\begin{eqnarray}
\label{SS}
t_{1/2}(u) & = &
q^{-\beta}t_0(u+\eta/2)\prod^{n}_{m=1}
\frac{\sin(u-v_m-\eta)}{\sin(u-v_m)}\\
\nonumber
&+&q^{ \beta}t_0(u-\eta/2)\prod^{n}_{m=1}
\frac{\sin(u-v_m+\eta)}{\sin(u-v_m)},
\end{eqnarray}
where $n=\frac{N}{2}-s$ is the number of the flipped spins in the sector
$S^z=s$, and $t_0(u)$ is the transfer matrix in the representation with
the spin $j=0$. The Bethe equations are equivalent to the conditions that
the transfer matrix eigenvalues determided by Eq.~(\ref{SS})
have no poles in the complex plane of the variable~$u$.
In the variables
\begin{equation}
Q(u) = \prod^{n}_{m=1} \sin(u - v_m ).
\end{equation}
Eq.~(\ref{SS}) is equivalent to the Baxter $T$-$Q$-equation
\begin{equation}
\label{TQ1}
t_{1/2}(u) Q(u) = q^{-\beta} t_0(u+\eta/2) Q(u-\eta) +
q^{\beta} t_0(u-\eta/2) Q(u+\eta).
\end{equation}
As shown in~\cite{BelStr}, it is convenient to interpret this equation
as a discrete realization of a second-order differential equations~\cite{BLZ}.
In additions to the functions $Q(u)$, this equation should then have a second
lineary independent solution $P(u)$  with the same eigenvalue of
the transfer matrix  $t(u)$. The transfer matrix eigenvalues can be expressed
through the eigenvalues of the Baxter operator~$Q(u)$ and the operator~$P(u)$
\begin{eqnarray}
\label{TRANSFERJ}
t_{j}(u) & = & q^{(2j-1)\beta}f(u-(j-1/2)\eta) \times \\
\nonumber
&\times &[ Q(u-(j+1/2)\eta)P(u+(j+1/2)\eta) -\\
\nonumber
&-& Q(u+(j+1/2)\eta)P(u-(j+1/2)\eta) ],
\end{eqnarray}
where the function $f(u)$ is quasiperiodic,
\begin{equation}
f(u+\eta) = q^{-2 \beta} f(u).
\end{equation}
This function is introduced for convenience (the function
$\tilde P(u) = f(u) P(u)$ could be introduced instead).
Because the functions $t_{j}(u)$ satisfy the same functional
fusion relation~(\ref{FUSION}), we conclude that these functions are
some eigenvalues of the corresponding transfer matrices $\hat{t}_{j}(u)$.

We now consider the case $q^{p+1}=-1$, where $p$ is a natural number.
For the transfer matrix in the representation with the spin $j=p/2$,
we have
\begin{equation}
t_{p/2}(u+\frac{\pi}{2}) = (-1)^{\beta}f(u)
[P(u+\pi) Q(u) - P(u) Q(u+\pi)].
\end{equation}
It then follows that if the functions $Q(u)$ and $P(u)$ satisfy the relation
\begin{equation}
\frac{P(u+\pi)}{P(u)} = \frac{Q(u+\pi)}{Q(u)},
\end{equation}
then the eigenvalue of the transfer matrix $t_{p/2}(u)$ is zero.

By analogy with the procedure in~\cite{BelStr}, we can express the function
$P(u)$ through the function $Q(u)$. We use the expansion
\begin{equation}
\frac{t_0(u)}{Q(u+\eta/2)Q(u-\eta/2)}=
R(u) + q^{\beta}\frac{A(u+\eta/2)}{Q(u+\eta/2)}
-q^{-\beta}\frac{A(u-\eta/2)}{Q(u-\eta/2)},
\end{equation}
where $R(u)$ is a trigonometric polynomial of the order $N-2 n$,
uniquely determined through the known trigonometric polynomials
$t_0(u)$ and $Q(u)$, and the order of the polinomial $A(u)$ is smaller
than $n$. Further, we use another expansion,
\begin{equation}
\label{RFF}
R(u) = q^{\beta}F(u+\eta/2) - q^{-\beta}F(u-\eta/2).
\end{equation}
It is now easy to verify that the function $P(u)$ is determined by
the expression
\begin{equation}
P(u) = \frac{1}{f(u)}(Q(u) F(u) + A(u)).
\end{equation}
This expression implies the following constraint imposed on the values
of the twist parameter $\beta$ in the boundary conditions of the $XXZ$ model:
\begin{equation}
\label{BETA}
(-1)^{2 \beta} = (-1)^N.
\end{equation}
This means that if the length of the spin chain of atoms is even,
then the number $\beta$ must be integer, and if the length is odd,
the number must be half-integer. No truncation of the functional
relations occurs for other values of $\beta$.
Representation~(\ref{RFF}) of the function $R(u)$ and $F(u)$
in terms of the functions $F(u)$ in which $F(u)$ behaves exactly as $R(u)$
under the argument shift $u \to u + \pi$ exists if the inequality
\begin{equation}
\label{USL}
\prod^{+s}_{m=-s} \sin(\pi \frac{m+\beta}{p+1}) \neq 0,
\end{equation}
holds, becase this product appears in the denominator if $F(u)$
is expanded in the Fourier series. The variable $m$ takes either
integer or half-integer values from $-s$ to $+s$. Depending on whether
the chain length is even or odd, the numbers $\beta$ and $s$ are either
both integer or both half-integer, and their sum or difference is
therefore always integer. Condition~(\ref{USL}) is satisfied if
\begin{equation}
\label{SPIN}
| s | < \min( \beta, \; p+1 - \beta) \mod (p+1).
\end{equation}
Hence, analyzing the second lineary independent solution of the Baxter
equation, we conclude that for the anisotropy parameter equal to the
root of unity, $q^{p+1}=-1$, the transfer matrix with the spin
$j=p/2$ in the sector $S^z=s$ has zero eigenvalues if
conditions~(\ref{BETA})~and~(\ref{SPIN}) are satisfied.

\section{Algebraic structure}
Above, we proved the existence of vectors such that functional
relations~(\ref{FUSION}) restricted to these vectors are truncated
and have form of a closed system of equations for the eigenvalues
of the transfer matrix $t_{1/2}(u)$. In this section, we find all
vectors from the space of states of the model that satisfy this
condition.

As mentioned above, the connection between the
energies of the chains with different twist parameters~(\ref{ENERG})
exists because of the presence of the quantum group
symmetry ${\rm U_q(sl(2))}$. The generators of this symmetry are
the operators $S^z$, $X$ and $X^{t}$, where
\begin{equation}
X= \sum_{n=1}^{N}q^{\frac12(\sigma^z_1+\ldots+\sigma^z_{n-1})}
   \sigma^{+}_n
q^{-\frac12 (\sigma^z_{n+1}+\ldots+\sigma^z_{N})}
\end{equation}
It is easy to verify that $X^{p+1}=0$ for $q^{p+1}=-1$.

The operator $X$ acting on a vector from the sector $S^z=s$
transforms this vector to a vector from sector $S^z=s+1$:
$[S^z, X]=X$. Acting with this operator on the Hamiltonian, we obtain
\begin{equation}
\label{VVV}
X \; H^{(\beta)}_{XXZ} - H^{(\beta-1)}_{XXZ} \; X
= (\ldots)(1-q^{2(S^z-\beta+1)}).
\end{equation}
In this formula and in Eq.~(\ref{VVV2}) below, the dots stand for a
finite number of operators whose explicit form is cumbersome and
irrelevant to this consideration. Formula~(\ref{VVV}) proves
relation~(\ref{ENERG}). A similar relation holds for the transfer matrix.

The six vertex model transfer matrix related to the $XXZ$ Hamiltonian
is expressed through the matrix elements of the monodromy matrix
$L(u)=R_N(u)\ldots R_1(u)$,
\begin{equation}
t_j(u) = \sum_{n=1}^{2j+1}q^{-2\beta(j+1-n)}L^n_n(u).
\end{equation}
The $R$-matrix in the representation $\frac12 \otimes j$ on
lattice site with the number $a$ has the form
\begin{equation}
R_a (u) =
\left(
\begin{array}{cccc}
\sin(u+(\frac{1}{2}+\hat{H}_a)\eta) & \sin(\eta) \hat{F}_a \\
\sin(\eta) \hat{E}_a & \sin(u+(\frac{1}{2}-\hat{H}_a)\eta) \\
\end{array}
\right).
\end{equation}
The operators $\hat{E}$, $\hat{F}$, and $\hat{H}$ have the matrix elements
\begin{equation}
\begin{array}{cccc}
\pi_j(\hat{H})^m_n = (j+1-n) \: \delta_{m,n},  & m,\: n = 1,\: 2,
\: \ldots, \: 2j+1. \\
\pi_j(\hat{E})^m_n = \omega_{m} \: \delta_{m, n-1}, &
\pi_j(\hat{F})^m_n = \omega_{n} \: \delta_{m-1, n}, \\
\end{array}
\end{equation}
where $\omega_{n} \; = \; \sqrt{[n]_q [2j+1-n]_q}$
In this formula, we use the standard notation
$[x]_q \; = \; (q^x-q^{-x})/(q-q^{-1})$.
In particular, the operator $H$ for $j=1/2$ is $\frac12 \sigma^z$,
and the operators $E$ and $F$ are respectively~$\sigma^{+}$ and~$\sigma^{-}$.
(i.e., the generators of the quantum algebra ${\rm U_q(sl(2))}$:
$[\hat{H}, \hat{E}] = \hat{E}$, $[\hat{H}, \hat{F}] = - \hat{F}$,
$[\hat{E}, \hat{F}] = [2 \hat{H}]_q$).

Using the Yang-Baxter equation
\begin{equation}
\label{YB}
(R_{\frac12 \otimes j})^{a n}_{b m}(u-v)
(L_{\frac12})^{b}_{c}(u) (L_{j})^{m}_{k}(v)=
(L_{j})^{n}_{m}(v) (L_{\frac12})^{a}_{b}(u)
(R_{\frac12 \otimes j})^{b m}_{c k}(u-v),
\end{equation}
as in~\cite{We}, we obtain the commutation relations between the matrix
elements of the monodromy matrix and the operators $S^z$ and $X$,
\begin{eqnarray}
\label{COMM}
q^{S^z} L^n_k &=& q^{n-k} L^n_k q^{S^z}, \\
\nonumber
X L^n_k(u) &=& q^{2(j+1) - n - k} L^n_k(u) X \\
\nonumber
&+& \omega_{k-1} e^{+iu} q^{j+1-n} \:  L^n_{k-1}(u) q^{-S^z} \\
\nonumber
&-& \omega_{n}\; e^{+iu}  q^{j+2-k} \:  L^{n+1}_k(u) q^{+S^z}.
\end{eqnarray}
To obtain commutation relations~(\ref{COMM}), we consider the
monodromy matrix $(L_{1/2})^n_k(u)$ as $u \to -i \infty$.
The matrix element $(L_{1/2})^1_2(u)$ is proportional to the
operator $X$. We then pass to the limit $u \to -i \infty$
in Yang-Baxter equation~(\ref{YB}). Commutation
relations~(\ref{COMM}) allow expressing the transfer matrix itself
through the operator $X$. The formula for the transfer matrix
with the spin $j=p/2$ in the sector $S^z=s$ is
(see Appendix~A)
\begin{eqnarray}
\label{TP}
t_{p/2}(u) & = & \sum_{r,k=0}^{p} X^r \hat{M}_{r k} X^k,
\end{eqnarray}
where
\begin{eqnarray}
\label{OBOZ}
\hat{M}_{r k}  & = & (-1)^r \omega^{-1} e^{-i p u}
q^{-2s-(2\beta+1-r-k)p/2-2k} \\
\nonumber
&\times&\sum_{n=0}^{p}q^{(2(\beta-s)+r-k)n} C^n_r C^{p-n}_k \\
\nonumber
&\times&\sum_{m=0}^{p-r-k} \lambda_{m,p-r-k}
(L_{p/2})(u)^{1}_{p-r-k-m} X^{p-r-k-m},
\end{eqnarray}
$\omega  =  \prod^{2j}_{k=1} \omega_k$,
the numbers $C^N_n$ are the  $q$-binomial coefficients
$$
C^N_n = \frac{1}{[n]_q!} [N]_q [N-1]_q \ldots [N+1-n]_q =
\frac{[N]_q!}{[n]_q![N-n]_q!},
$$
and the variables $\lambda_{m, p-r-k}$ are determined by the formula
$$
\lambda_{m, m'} = e^{imu}q^{p/2+1+s+(m'-m)(m'+s-3p/2-1)}
C^{m'}_m  \prod_{l=1}^{m}
\omega_{m'-l}.
$$
As can be seen from formula~(\ref{TP}), the transfer matrix $t_{p/2}(u)$
is expressed through of monomials of the form $X^r \hat{M}_{r k} X^k$.

We now investigate the numbers $r$ and $k$ for which the operator
$\hat{M}_{r k}$ is nonzero in the case where $q^{p+1}=-1$.
The operator $\hat{M}_{r k}$ is proportional to the sum of the
$q$-binomial coefficients
\begin{eqnarray}
\label{SUMMA}
f_{r k} & \equiv & \sum^p_{n=0} C^n_{r} C^{p-n}_{k} q^{n (2 l +r-k)},
\end{eqnarray}
where $l=\beta-s$. Unfortunately, we failed to obtain an expression
for the coefficients $f_{r k}$ in a closed form. We therefore analyzed
sum~(\ref{SUMMA}) numerically. We verified the hypothesis that
for $q^{p+1}=-1$, the coefficients $f_{r k}$ are nonzero only in
the closed domain $r+k \le p$ and ($r>p-l$ or $k>l-1$).
$$
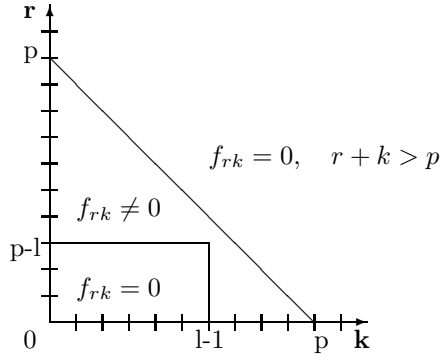
\begin{picture}(130,130)
\put(10,10){\vector(1,0){120}}
\put(10,10){\vector(0,1){120}}
\put(110,10){\line(-1,1){100}}
\put(70,10){\line(0,1){30}}
\put(10,40){\line(1,0){60}}
\put(125,0){{\bf k}}
\put(0,125){{\bf r}}
\put(110,0){p}
\put(0,110){p}
\put(65,0){l-1}
\put(-5,35){p-l}
\put(0,0){0}
\put(70,70){$f_{r k} = 0, \quad r+k>p$}
\put(20,20){$f_{r k} = 0$}
\put(20,50){$f_{r k} \neq 0$}

\put(20,7){\line(0,1){6}}
\put(30,7){\line(0,1){6}}
\put(40,7){\line(0,1){6}}
\put(50,7){\line(0,1){6}}
\put(60,7){\line(0,1){6}}
\put(70,7){\line(0,1){6}}
\put(80,7){\line(0,1){6}}
\put(90,7){\line(0,1){6}}
\put(100,7){\line(0,1){6}}
\put(110,7){\line(0,1){6}}
\put(120,7){\line(0,1){6}}

\put(7,20){\line(1,0){6}}
\put(7,30){\line(1,0){6}}
\put(7,40){\line(1,0){6}}
\put(7,50){\line(1,0){6}}
\put(7,60){\line(1,0){6}}
\put(7,70){\line(1,0){6}}
\put(7,80){\line(1,0){6}}
\put(7,90){\line(1,0){6}}
\put(7,100){\line(1,0){6}}
\put(7,110){\line(1,0){6}}
\put(7,120){\line(1,0){6}}
\end{picture}
$$
Hence, the transfer matrix $t_{p/2}(u)$ can be expressed as
\begin{equation}
\label{VVV2}
t_{p/2}(u) = X^{p+1-l}(\ldots) \; + \; (\ldots) X^{l}
\end{equation}
and therefore vanishes on the cohomologies
\begin{equation}
V_{p, l} = \Ker X^{l}/\im X^{p+1-l}.
\end{equation}
This means that $t_{p/2}(u)$ has zero eigenvalues on the eigenvectors
$v$ form the sector $S^z \; = \; s$ that are annihilated by the
operator $X^l$,
\begin{equation}
X^l \; v \; = \; 0,
\end{equation}
and cannot be expressed in the form $X^{p+1-l} \chi$, where $\chi$
is again vector,
\begin{equation}
v \; \neq \; X^{p+1-(\beta-s)} \chi.
\end{equation}
We have thus identified all vectors on vich the transfer
matrix $t_{p/2}(u)$ has zero eigenvalues and, as mentioned above,
recursive system~(\ref{FUSION}) takes the form of a closed system
of equations on the eigenvalues of the transfer matrix $t_{1/2}(u)$.

An example illustrating how the truncated functional relations
can be used is given in Appendix~B. Specifically, the eigenvalues
of the transfer matrix and the Hamiltonian are obtained there for
the case~$q^4=-1$.

\section{Appendix A}
Formula~(\ref{TP}) can be directly derived by using commutation
relations~(\ref{COMM}), but this way is rather wearisome.
The difficulties are connected with the action of the operator
$X$ on $L^n_k(u)$ resulting in the appearance of two terms in
the right-hand side, namely, $L^{n}_{k-1}(u)$ and $L^{n+1}_{k}(u)$.
The calculation, however, can be simplified. We note that the
commutation relations for $X$ and $L^n_k(u)$ are formally equivalent
to the commutation relations for $X$ and the auxiliary operator
$\psi^n(u) q^{-n S^z} \psi_m(u) q^{-m S^z}$, where the auxiliary
objects $\psi$ commute with $X$ according to the rule
\begin{eqnarray}
\label{PSICOMM}
X \psi_j^n -q^{j+1-2n}\psi_j^nX & = & -\omega_n e^{i u}
q \psi_j^{n+1} \\
\nonumber
X \psi_n^j -q^{j+1-2n}\psi_n^jX & = & \omega_{n-1} e^{i u}
\psi_{n-1}^j \\
\nonumber
q^{S^z} \psi^n_j & = & q^{n-j-1} \psi^n_j q^{S^z}\\
\nonumber
q^{S^z} \psi^j_n & = & q^{j+1-n} \psi^j_n q^{S^z}
\end{eqnarray}
With this definition of the auxiliary objects $\psi$,
the correspondence rule
\begin{eqnarray}
\label{LPSI}
(L_j)^n_k(u) & \sim & \psi^n_j(u) \; q^{-n S^z} \;
\psi_k^j(u) \; q^{-k S^z},
\end{eqnarray}
can be formulated. The symbol $\sim$ means that the commutation
relations for the variables $X$ and $F(L^n_k)$, where $F$ is an
arbitrary function, are exactly the same as those for $X$ and
$F(\psi^n(u)  q^{-n S^z}  \psi_k(u) q^{-k S^z})$.
The commutation relations for the latter variables, however, are
easier to obtain because commutation relations~(\ref{PSICOMM})
are simpler than relations~(\ref{COMM}). Hence, the symbol $\sim$
in the Eq.~(\ref{LPSI}) should be interpretted as "{\it transforms as}".
Hereinafter, we use the equality sign instead of the symbol
"$\sim$" for simplicity.

Relations~(\ref{PSICOMM}) can be rewritten as
\begin{eqnarray}
\label{PSICOMMREC}
\psi_j^{n}  & = & -(\omega_{n-1} q)^{-1} e^{-iu}
( X \psi_j^{n-1} -q^{j+1-2(n-1)}\psi_j^{n-1} X)\\
\nonumber
\psi_{n}^j & = & (\omega_{n+1})^{-1} e^{-iu}
( X \psi_{n+1}^j -q^{j+1-2(n+1)}\psi_{n+1}^j X).
\end{eqnarray}
These simple recursive relations allow expressing $\psi^n$
through $\psi^1$ and $\psi_n$ through $\psi_{2j+1}$.
The solution of thes equations have the forms
\begin{eqnarray}
\label{YAPSI}
\psi_j^n&=&a_n \sum_{r=0}^{n-1}(-1)^r q^{r(j+1-n)}
C^{n-1}_r X^{n-1-r} \psi^1 X^r \\
\nonumber
\psi_n^j&=&b_n \sum_{k=0}^{2j+1-n}(-1)^kq^{-k(j+1+n)}
C^{2j+1-n}_{k}X^{2j+1-n-k}\psi_{2j+1}X^{k},
\end{eqnarray}
where the coefficients are
\begin{eqnarray}
a_n &= &\prod^{n-1}_{k=1}(-e^{-iu}q^{-1}\omega^{-1}_{n})\\
\nonumber
b_n &= &\prod^{2j+1-n}_{k=1}(e^{-iu} \omega^{-1}_{2j+1-k}).
\end{eqnarray}
The transfer matrix is
\begin{eqnarray}
t_{j}(u) & = & \sum_{n=1}^{2j+1}q^{-2\beta(j+1-n)}(L_j)^n_n(u) \\
\nonumber
& = & \sum_{n=1}^{2j+1}q^{-2\beta(j+1-n)}
\psi^n_j(u) q^{-n S^z}  \psi_n^j(u) q^{-n S^z}.
\end{eqnarray}
According to formulas~(\ref{YAPSI}), we obtain
\begin{eqnarray}
\nonumber
t_{p/2}(u) & = & \omega^{-1} e^{-i p u} q^{-2s-p(2\beta+1)/2}
\sum^p_{n=0} \sum^n_{r=0} \sum^{p-n}_{k=0}
(-1)^r q^{-p r/2 - (p/2+2)k}\\
& \times  &
q^{n(2(\beta-s)+r-k)} C^n_r C^{p-n}_k X^r
\left( \psi^1 X^{p-r-k}\psi_{p+1} \right) X^k.
\end{eqnarray}
It remains to remove the auxiliary operators $\psi^1$
and $\psi_{p+1}$ that participate in the formula in the
form $\psi^1 X^{p-r-k}\psi_{p+1}$. To do this, we commute the operators
$X^{p-r-k}$ and $\psi_{p+1}$. Using~(\ref{PSICOMM}), it is easy to prove
by the induction that
\begin{eqnarray}
\nonumber
\psi^A_j X^M \psi^j_B & = & \sum_{m=0}^{M}e^{imu}
q^{(M-m)(j+1+m-2B)+A(j+1-B+m)}C^M_m \\
& \times & \left( \prod_{l=1}^{m}\omega_{B-l} \right)
L(u)^A_{B-m} q^{(A+B-m)S^z}X^{M-m}.
\end{eqnarray}
Using this formula and taking the relation $[S^z, X]=X$ into account,
we obtain expression~(\ref{TP}) for~$t_{p/2}(u)$. Only $X$ and $L$
participate in the final formula; the auxiliary objects $\psi$
are no longer needed.

\section{Appendix B}
We obtain the eigenvalues of the transfer matrix and the Hamiltonian for
$q^4 = -1$. We have
\begin{eqnarray}
t_{3/2}(u) & = & 0,\\
\nonumber
t_{1-j}(u) & = & -(-1)^{\frac{N}{2}+\beta-s} t_j(u+\pi/2).
\end{eqnarray}
To simplify the calculation, we introduce the function
$S_N(u) \equiv (-2)^{N/2} t_{1/2}(u-\eta/2)$.
The dunctional relations become
\begin{equation}
\label{FFF}
S_N(u+\frac{\pi}{8})S_N(u-\frac{\pi}{8}) = \cos^N(2u) -
(-1)^{-\frac{N}{2}+l} \sin^N (2u).
\end{equation}
Equation~(\ref{FFF}) contains the numbers $N$ and $l=\beta-s$.
We obtain four different equations, depending on whether these
numbers are even or odd. Correspondigly, we obtain four different solutions:
\begin{eqnarray}
S^{ \{ n \} }_{4 M}(u) & = & \prod^{M}_{m=1}(1 + (-1)^{n_m}
\cos(\pi\frac{2m-1}{4M})\cos(4u)), \\
\nonumber
S^{ \{ n \} }_{4 M + 2}(u) & = & \sin(2u) \sqrt{2} \prod^{M}_{m=1}
(1 + (-1)^{n_m} \cos(\pi\frac{2m}{4M+2})\cos(4u)),
\end{eqnarray}
$$
S^{ \{ n \} }_{4 M -1}(u)  =  e^{- i u}(-i)^{M-1/2} \times
$$
$$
\prod^{2 M -1}_{m=1} (e^{(-1)^{n_m}
\frac{ i \pi m}{4M-1}}\sin(2u+\frac{\pi}{4})
-i e^{-(-1)^{n_m}\frac{ i \pi m}{4M-1}}\sin(2u-\frac{\pi}{4})),
$$
$$
S^{ \{ n \} }_{4 M +1}(u)  =  e^{- i u}(-i)^{M} \times
$$
$$
\prod^{2 M}_{m=1}
(e^{(-1)^{n_m}\frac{i \pi m}{4M+1}}\sin(2u+\frac{\pi}{4})
- i e^{-(-1)^{n_m}\frac{i \pi m}{4M+1}}\sin(2u-\frac{\pi}{4})),
$$
In this formula, the set $\{ n_m \}$ is any set of numbers equal
to either $0$ or $1$, for example, $\{ 0, 1, 0, \ldots, 0, 1 \}$.
We interpret them as the fermion occupation numbers.
The functions written above satisfy the equation
\begin{equation}
S_{4 M + \delta}(u+\frac{\pi}{8})S_{4 M + \delta}(u-\frac{\pi}{8})=
\cos^{4 M + \delta}(2 u) + i^{ - \delta}\sin^{4 M + \delta}(2u).
\end{equation}
Hence, we obtained the solutions only for the case where
$\beta-s$ is an odd number. For the energy eigenvalues, we obtain
\begin{eqnarray}
E^{\{n\}}_{4M} & = & -\frac{4M}{2\sqrt{2}} - \frac{4}{\sqrt{2}}
\sum^{M}_{m=1}(-1)^{n_m}\cos(\pi \frac{2m-1}{4M}), \\
\nonumber
E^{\{n\}}_{4M+2} & = & -\frac{4M+2}{2\sqrt{2}} +
\frac{2}{\sqrt{2}} \left( 1
- 2 \sum^{M}_{m=1}(-1)^{n_m}\cos(\pi \frac{m}{2M+1}) \right), \\
\nonumber
E^{\{n\}}_{4M-1} & = & -\frac{4M-1}{2\sqrt{2}} -
\frac{1}{\sqrt{2}} \left( i
+ 2 i \sum^{2M-1}_{m=1}\exp(\frac{-2i(-1)^{n_m}\pi m}{4M-1})
\right), \\
\nonumber
E^{\{n\}}_{4M+1} & = & -\frac{4M+1}{2\sqrt{2}} -
\frac{1}{\sqrt{2}} \left( i
+ 2 i \sum^{2M}_{m=1}\exp(\frac{-2i(-1)^{n_m}\pi m}{4M+1}) \right).
\end{eqnarray}
Although the imaginary unit explicitly participates in the last
two formulas, the energy $E$ is real in all the cases because
the imaginary parts cancel each other in the sum. For spin chains whose
length are even, $N=4M$ and $N=4M+2$, we obtain $2^M$ energy levels.
For spin chains whose length are odd, $N=4M-1$ and $N=4M+1$, we
respectively obtain $2^{2M-1}$ and $2^{2M}$ eigevalues.

\paragraph{Acknowledgments}
The authors thank M. Yu. Lashkevich for the useful comments and fruitful
discussions. This work is partially supported by RFBR 01-02-16686
and 00-15-96579, CRDF RP1-2254 and INTAS-00-00055.

\end{document}